# Ultralow Threshold, Single-Mode InGaAs/GaAs Multi-Quantum Disk Nanowire Lasers

Xutao Zhang[†], Ruixuan Yi[†], Nikita Gagrani[ξ], Ziyuan Li[ξ], Fanlu Zhang[ξ], Xuetao Gan[†*], Xiaomei Yao[ψ§], Xiaoming Yuan[#], Naiyin Wang[ξ], Jianlin Zhao[†], Pingping Chen[ψ§*], Wei Lu[ψ§α*], Lan Fu[ξ@], Hark Hoe Tan[ξ@], and Chennupati Jagadish[ξ@]

[†]Key Laboratory of Light Field Manipulation and Information Acquisition, Ministry of Industry and Information Technology, and Shaanxi Key Laboratory of Optical Information Technology, School of Physical Science and Technology, Northwestern Polytechnical University, Xi'an 710129, China

[ξ]Department of Electronic Materials Engineering, Research School of Physics, The Australian National University, Canberra, ACT 2601, Australia.

[ψ]State Key Laboratory for Infrared Physics, Shanghai Institute of Technical Physics, Chinese Academy of Sciences, 500 Yutian Road, Shanghai 200083, China

[§]University of Chinese Academy of Sciences, 19 Yuquan Road, Beijing 100049, China

[#]Hunan Key Laboratory of Nanophotonics and Devices, School of Physics and Electronics, Central South University, 932 South Lushan Road, Changsha, Hunan 410083, P. R. China.

[α]School of Physical Science and Technology, ShanghaiTech University, 393 Middle Huaxia Road, Pudong District, Shanghai 201210, China.

[@]ARC Centre of Excellence for Transformative Meta-Optical Systems, Research School of Physics, The Australian National University, Canberra, ACT 2601, Australia.

**ABSTRACT:** We present single-mode nanowire (NW) lasers with ultralow threshold in the near-infrared spectral range. To ensure the single-mode operation, the NW diameter and length are reduced specifically to minimize the longitudinal and transverse modes of the NW cavity. Increased optical losses and reduced gain volume by the dimension reduction are compensated by excellent NW morphology and InGaAs/GaAs multi-quantum disks. At 5 K, a threshold low as 1.6 μJ/cm$^2$ per pulse is achieved with a resulting quality factor exceeding 6400. By further passivating the NW with an AlGaAs shell to suppress surface non-radiative recombination, single-mode lasing operation is obtained with a threshold of only 48 μJ/cm$^2$ per pulse at room temperature with a high characteristic temperature of 223 K and power output of ~ 0.9 μW. These single-mode, ultralow threshold, high power output NW lasers are promising for the development of near-infrared nanoscale coherent light sources for integrated photonic circuits, sensing, and spectroscopy.

Nanowire (NW) lasers with their ultracompact footprint, ease of integration, and low energy consumption are of great interests for future integrated photonic devices in optical interconnects and super-resolution imaging.[1-7] Due to its natural geometry, the NW itself forms an optical cavity and by using materials such as III-V semiconductors as the NW, gain can also be achieved simultaneously. The NW's two end facets function as the mirrors of a Fabry-Perot (F-P) cavity.[8-10] Unfortunately, despite its physical dimension, there are normally multiple longitudinal and transverse modes of NW-based F-P cavity that overlap with the NW gain spectrum.[11] It results in the multimode operation of the NW lasers, which is undesirable in many applications and would hinder the future applications of NW lasers. For instance, multimode operation could cause time-domain pulse broadening and signal crosstalk in optical communications, limiting the bandwidth.[12-14]

In past decades, a variety of strategies have been developed to achieve single-mode NW lasers. Coupling multiple NW cavities could increase the free spectral range (FSR) of the resonant modes using the Vernier effect,[15, 16] thereby allowing only one mode to overlap with the gain curve for the single-mode lasing. However, this would involve awkward manipulations of NWs down to the nm-scale accuracy in order to align the NWs. By patterning the NW with a periodic nanostructure or placing the NW on a grating, single lasing mode could be achieved *via* the distributed feedback effect.[12, 17] However, this implementation involves the complicated nanofabrication processes on the NW, which can degrade the material properties. The number of resonant modes in NW F-P cavity can also be reduced by reducing the NW diameter and length, and single-mode NW lasers have been reported.[11, 18] However, reduction in NW dimension increases the optical loss significantly considerably, especially in the near-infrared range.[19-21] In addition, the active gain volume is reduced by reducing the NW dimension, leading to higher lasing threshold.

In this work, we report single-mode NW lasers with ultralow threshold at the wavelengths around 950 nm. Through computational methods, the NW length and diameter are optimized to ~ 2.2 μm and ~ 200 nm, respectively, to support single-mode lasing. This is the smallest dimension reported to date for NW lasers. To overcome the large optical loss and small gain volume, multiple GaAs/InGaAs quantum disks (QDs) are grown in the central region of the GaAs NW to provide sufficient optical gains. At 5 K, the lasing threshold is low as 1.6 µJ/cm$^2$ per pulse with a *Q-factor* exceeding 6400. By further passivating the NW with an AlGaAs shell uniformly to suppress surface non-radiative recombination, room temperature single-mode lasing operation is obtained with a low threshold of 48 µJ/cm$^2$ per pulse and a power output ~ 0.9 µW. The merits of single-mode, ultralow threshold, and room temperature operation make near-infrared NW laser a promising nanoscale coherent light source for integrated photonic circuits, sensing and spectroscopy.

**RESULTS AND DISCUSSION**

The inserted GaAs/InGaAs QDs in the GaAs NW have a spontaneous emission around 1 µm in wavelength with a linewidth of 50 nm (see Figure S1). To realize single-mode lasing, the geometry of the NW is designed with the assistance of mode simulations using the two dimensional finite difference eigenmode (FDE) solver. The number of transverse and longitudinal optical modes of the NW cavity is minimized around the gain spectral range. In the simulation, a pure GaAs NW lying on a SiO$_2$ substrate (index ~ 1.45) is employed considering the refractive indices of GaAs and InGaAs are similar. Due to the large refractive index of GaAs, the NW supports multiple transverse modes, including HE$_{11}$, HE$_{21}$, EH$_{11}$, TE$_{01}$, TM$_{01}$, *etc*. The effective refractive indices of these modes are calculated for NWs of different diameters, as shown in Figure 1a. Here, the modes with effective refractive indices larger than 1.45 are plotted considering the underlying SiO$_2$ substrate. Note, because the SiO$_2$ substrate breaks the symmetry of the NW modes, the degenerate modes are separated slightly, such as HE$_{11a}$ and HE$_{11b}$, HE$_{21a}$ and HE$_{21b}$. When the NW diameter is larger than 165 nm, the effective refractive index of the HE$_{11a}$ mode is higher than 1.45, which could support a resonance mode. As the diameter increases, the number of supported resonance modes in the NW cavity also increases. When the diameter of the NW is in the range of 165 - 230 nm, only two transverse modes HE$_{11a}$ and HE$_{11b}$ are supported in the cavity. This is the preferred range of the NW diameter for realizing single transverse mode lasing. The mode spacing ($\Delta\lambda$) can be obtained through:

$$\Delta\lambda = \frac{\lambda^2}{2L}\left(n_{eff} - \lambda\frac{dn_{eff}}{d\lambda}\right)^{-1}$$

in which λ is center wavelength (1000 nm), $L$ is the length of NW and $n_{eff}$ is the effective refractive index of guided mode. According to the definition of group index $n_g$, the mode spacing could also be expressed as follows:

$$\Delta\lambda = \frac{\lambda^2}{2L}n_g^{-1}$$

By simulation in Lumerical MODE analysis, the group index of $HE_{11a}$ and $HE_{11b}$ are calculated to be 5.60 and 5.99, respectively. Figure 1b shows the change of the mode spacings of $HE_{11a}$ and $HE_{11b}$ modes with NW length for a fixed diameter of 200 nm. When the length of NW equals to 3.3 μm, the mode spacings of $HE_{11a}$ and $HE_{11b}$ are approximate 30 and 25 nm. Moreover, the mode spacings increase as the length of NW decreases. For a gain curve with a spectral width of ~ 50 nm (in the case of our InGaAs/GaAs QDs), then there only one mode within this bandwidth when the length is less than 3.3 μm (see the inset of Figure 1b). Therefore single longitudinal mode lasing can be expected.

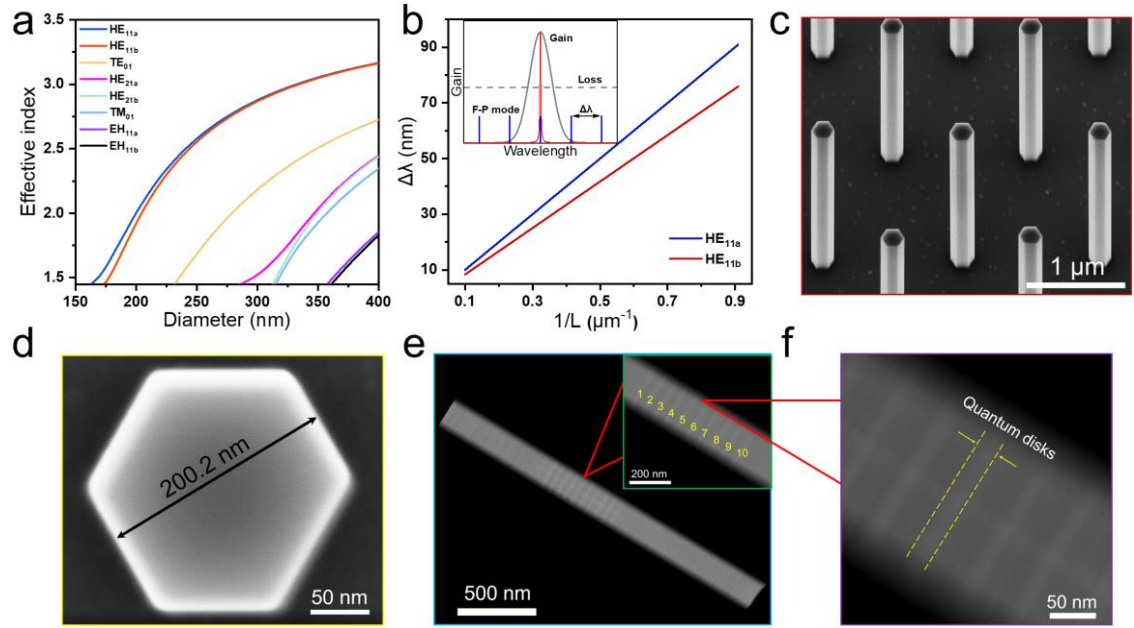

Figure 1. Design and structural characterization of the NWs. (a) Modal effective refractive index for different modes as a function of the NW diameter. (b) Modal optical mode spacing as a function of reciprocal of NW length with a diameter of 200 nm. The inset plot shows the QD gain spectrum, the expected NW laser output and the various

F-P modes sustained by the NW cavity. (c) SEM image at 30° tilt angle of NW array standing on the GaAs(111)B substrate (the diameter of the mask opening is 40 nm). (d) Top view of a vertical NW, showing a diameter of about 200 nm. (e) STEM image of a transferred NW, showing that the QDs are located about 1.2 μm from the bottom of the NW. The inset in the upper right corner shows that the NW contains 10 InGaAs QDs. (f) Magnified view showing that the thickness of the QD is approximately 19 nm.

With the optimized design, the GaAs NWs with ten $In_{0.2}Ga_{0.8}As$/GaAs QDs are grown by selective area metal-organic vapor phase epitaxy (SA-MOVPE) to the required dimensions (see Figure S2 and the Methods details). To obtain high optical efficiency, the composition and thickness of these ten QDs should be as uniform as possible. Figure 1c shows the 30° tilted scanning electron microscope (SEM) image of the NW array on the growth substrate. The NWs are hexagonal prisms with a length of 2.2 μm, which are uniform and smooth from the bottom to the top without any tapering. These features are important to maintain a stable guided mode along the NW to obtain a high $Q$ factor in the optical cavity. The top view shown in Figure 1d indicates the NWs have a diameter of 200 nm, which can only support the $HE_{11a}$ and $HE_{11b}$ modes. High-angle annular dark field scanning transmission electron microscopy (HAADF-STEM) analyses show that multi-QD region (brighter regions correspond to InGaAs QDs), is located about 1.2 μm above the bottom of the NW. The magnified view of the QDs region in the inset of Figure 1e shows the ten InGaAs QDs, separated by GaAs barriers. Figure 1f further shows a higher magnification image of the QDs region where the average thickness of the QDs and barrier are ~ 19 and ~ 35 nm, respectively. These relatively thick QDs (and 10 QDs) are designed and grown to provide sufficient gain.[22] Through statistical analysis, we found that the QDs are quite uniform in thickness, an important criterion to obtain high gain and realize low lasing threshold.

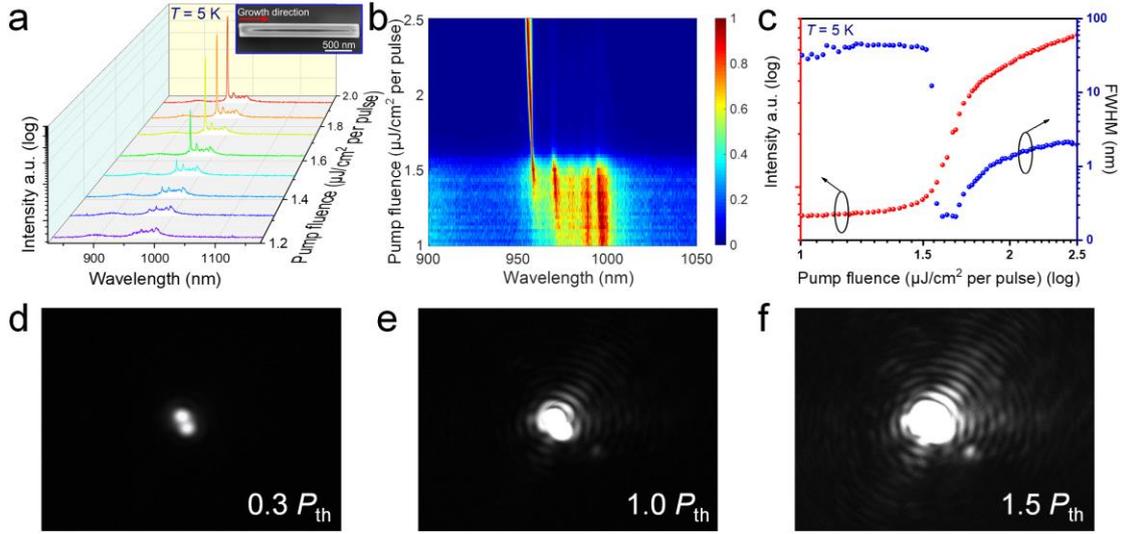

Figure 2. Lasing characterization of a GaAs/InGaAs multi-QDs NW at 5 K. (a) Emission spectra at different pump fluences. (The y-axis is plotted on a logarithmic scale). The inset in the upper right corner is an SEM image of single NW transferred onto 285 nm-thick $SiO_2$/Si substrate. (b) Spectral map of normalized emission intensity *vs* pump fluence. (c) Lasing emission intensity (red) and the corresponding FWHM of the spectrum (blue) *vs* pump fluence plotted on a logarithmic-logarithmic scale. (d-f) Optical images from single NW under a pump fluence of 0.3 $P_{th}$ (d), 1.0 $P_{th}$ (e) and 1.5 $P_{th}$ (f).

To evaluate the lasing properties of these NWs, individual NWs were mechanically transferred to a 285 nm-thick $SiO_2$/Si substrate by using a tungsten wire probe, as shown in the upper right corner of Figure 2a. Optical pumping of individual NWs was conducted at 5 K in confocal micro-photoluminescence (μ-PL) system (see the schematic diagram of the experimental setup in Figure S3). Figure 2a shows the emission spectrum from the NW with increasing pump fluence. A broad PL spectrum of about 50 nm in width centered at 975 nm appears at low pump fluences. As the excitation pump fluence is increased, the PL spectrum blue shifts due to band filling effects.[20, 23] When the excitation pump fluence is increased to 1.6 μJ/cm$^2$ per pulse, a narrow peak appeared at 959 nm. With further increase in pump fluence, the intensity of this narrow peak increases significantly and dominates the entire emission spectrum with clamping spontaneous emission becoming apparent. Figure 2b shows the spectral map of the NW with increasing pump fluence. A narrow single-mode lasing emission can be clearly observed beyond a fluence of 1.6 µJ/cm² per pulse. The light output *vs* light input curve (L-L curve) plotted on a log-log scale in Figure 2c exhibits a typical "S"-shape, a clear indication of the transition from spontaneous emission to amplified spontaneous emission to stimulated emission, which increasing fluence. The full width at half maximum (FWHM) of the lasing peak *vs* excitation fluence is also shown in Figure 2c. In spontaneous emission region, it increases gradually with pump fluence but drop suddenly above the lasing threshold. A minimum linewidth of 0.15 nm is achieved (Figure S4), but increases with further increase pump fluence due to thermal broadening effect. The *Q-factor* is calculated to be 6400 using the formula[24] $Q = \lambda/\Delta\lambda$ ($\lambda$ is the wavelength of lasing peak). The high *Q-factor* is a result of the highly uniform and regular NW cavity, resulting in a very low lasing threshold $P_{th}$ of ~1.6 µJ/cm² per pulse, which is much lower than previously reported for NW lasers.[25-29] Furthermore, the threshold can be further reduced by increasing the number of QDs. It is worth mentioning that the free spectral range is ~ 40 nm for our NWs with a length of 2.2 µm, so only one optical longitudinal mode can be supported in the 50 nm gain curve of the InGaAs QDs.

Optical imaging is used to further verify the coherence of stimulated emission from the NW laser, as shown in Figures 2d-f. When the pump fluence intensity is 0.3 $P_{th}$, two bright spots can be seen on the two end facets of the NW (Figure 2d), indicating the NW forms a good waveguide. At threshold, emission from the two end facets start to interact coherently, forming fringes as seen in Figure 2e. These coherent fringes become more pronounced beyond lasing threshold as shown in Figure 2e at a fluence of 1.5 $P_{th}$, conclusively showing that the NW is lasing.

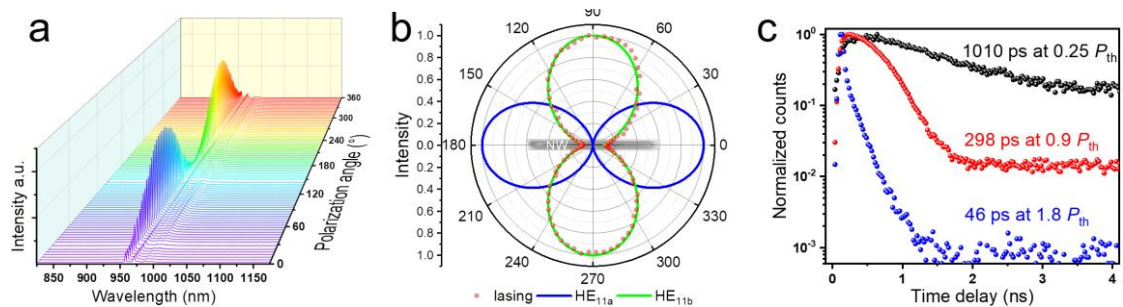

Figure 3. (a) Polarization-dependent emission spectra of a GaAs/InGaAs multi-QDs NW under a pump fluence of 1.3 $P_{th}$ at 5 K. (b) Polarization dependence of the lasing intensity (red data points) from experiment and different guides modes ($HE_{11a}$ blue curve, $HE_{11b}$ green curve) from 3D FDTD simulations, which are plotted in polar coordinates relative to the orientation of NW. (c) Normalized time-resolved emission decays from the NW under a pump fluence of 0.25 $P_{th}$ (d), 0.9 $P_{th}$ (e) and 1.8 $P_{th}$ (f).

To determine the transverse mode of lasing, we evaluate the polarization dependence of the emission spectrum by changing the angle of the linear polarizer placed along in the signal collection path, which is shown in Figure 3a. The lasing peak intensities under different polarization angles are then extracted from Figure 3a and plotted in polar coordinates relative to the orientation of NW, as shown in Figure 3b. The lasing intensity shows strong linear polarization dependence with a high extinction ratio $\rho = (I_{\parallel} - I_{\perp})/(I_{\parallel} + I_{\perp})$ of 84%, in which the polarization direction is perpendicular to NW axis caused by the characteristic of lasing transverse mode. 3D finite difference time domain (FDTD) simulation[30] is further used to understand the polarization dependence of the $HE_{11a}$/$HE_{11b}$ modes (these two transverse modes are only supported in these NWs with a diameter of 200 nm), as shown in Figure 3b (see supporting information for details). The experimentally measured polarization direction is the same as $HE_{11b}$ mode, and perpendicular to $HE_{11a}$ mode. Furthermore, the experimentally measured polarization extinction ratio also matches well with the simulation results of $HE_{11b}$ mode, suggesting that the lasing mode is $HE_{11b}$.

Figure 3c shows that the minority carrier lifetime at 5 K shortens with pump fluence, which further confirms lasing from the NWs. At low excitation (0.25 $P_{th}$), most of carriers are consumed by spontaneous emission and such a long lifetime of 1010 ps even in NWs without surface passivation indicates the surface recombination is suppressed at low temperature. Above threshold, the carrier lifetime significantly shortens to 46 ps, at 1.8 $P_{th}$, indicating that stimulated emission now dominates the carrier recombination process. Noted that 46 ps has reached temporal resolution of our system (~ 50 ps), the actual carrier lifetime after stimulated emission could be shorter. In addition, the morphology and quality of each NW in the array are very good, resulting in a high yield and performance (see Figure S5).

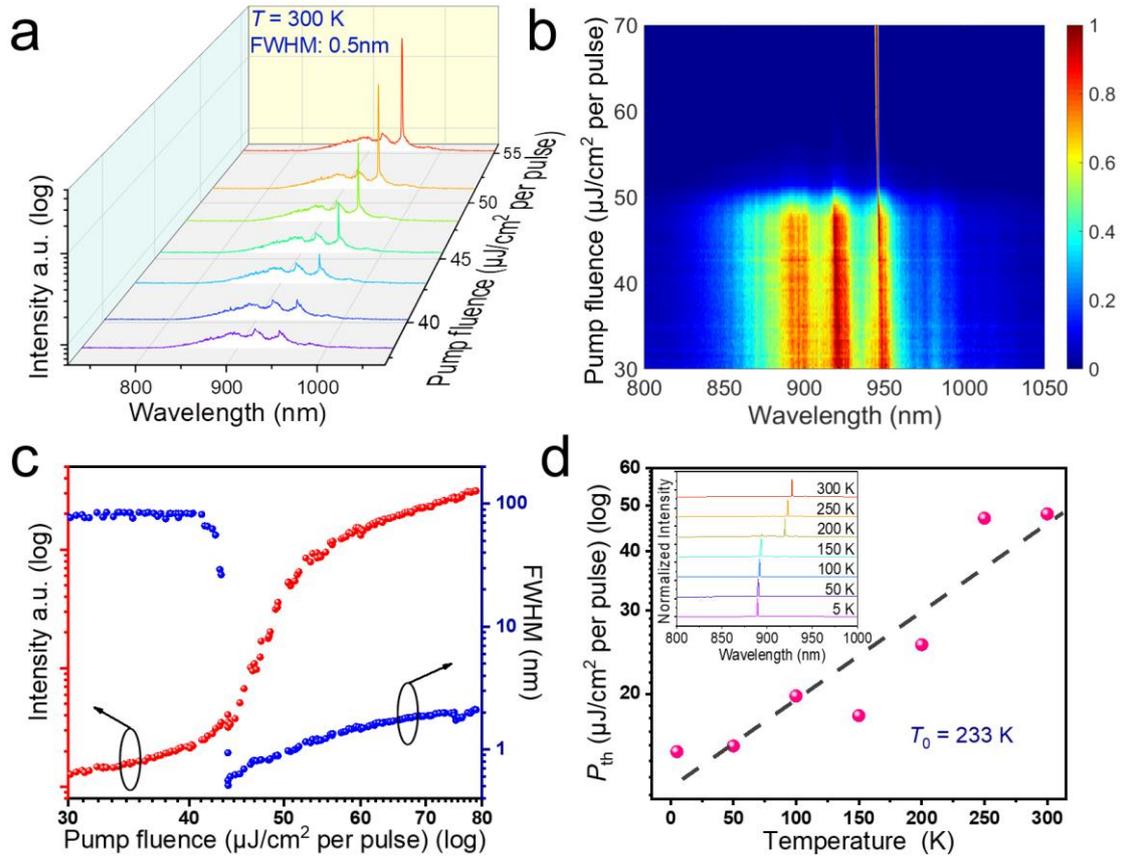

Figure 4. Lasing characterization of a GaAs/InGaAs/AlGaAs multi-QDs NW at 300 K. (a) Emission spectra at different pump fluences plotted on a logarithmic scale in the y-axis. (b) Spectral map of normalized emission intensity *vs* pump fluence. (c) Lasing emission intensity (red) and the corresponding FWHM of the spectrum (blue) *vs* pump fluence plotted on a logarithmic-logarithmic scale. (d) Threshold pump fluence *vs* temperature. Fitting the experimental data points yields a characteristic temperature of 233 K. The inset shows the emission spectra at various temperatures pumped at 1.1 $P_{th}$.

Due to the high *Q-factor* and optical gain of these NWs, lasing operation can be sustained up to a temperature of 180 K (see Figure S7). However, to further increase the operating temperature, an AlGaAs shell followed GaAs cap were grown on the NWs to passivate and suppress surface non-radiative recombination. Figures 4a-b and Figure S8 show the emission spectra at different pump fluences. The single mode narrow lasing peak centered at 945.6 nm generates at a pump fluence of 48 µJ/cm$^2$ per pulse has an extremely narrow linewidth about 0.5 nm at 300 K (see Figure 4c). Figure 4d shows lasing threshold fluence $P_{th}$ *versus* temperature from 5 to 300 K for the core-shell NW.

The $P_{th} \sim e^{T/T_0}$ is used to fit the data, resulting in $T_0 = 233$ K. Such a high characteristic temperature $T_0$ indicates strong quantum confinement of carriers in the QDs. The lasing spectra at various temperatures are shown in the inset of Figure 4d. From 5 to 150 K, the lasing peak shows a slight redshift. The slight redshift of lasing peak is caused by the change of refractive index with temperature. However, at 150 K, a new peak at 918 nm emerges (see also Figure S8) which then continues to redshift slightly with temperature. The emergence of new lasing peak from different transverse modes could be ascribed to the redshift of gain spectrum (see the details in Supporting information), with temperature increasing.[31]

Table 1. Comparison of cavity length, threshold, lasing peak and *Q-factor* of various NW lasers operating in the near-infrared region.

| Material | Cavity length/μm | Threshold/μJ cm$^{-2}$ per pulse | Lasing peak/nm | Q-factor | Temperature | References |
|---|---|---|---|---|---|---|
| InAs NW | 14 | 50 | 2400-2700 | 520 | 4 K | 19 |
| GaAs/ GaNAs/GaAs dilute nitride NW | 6.3 | 16 | 1000 | 625 | 5 K | 25 |
| PbS NW | 10-15 | - | 3-4 μm | 428 | 180 K | 23 |
| InP/InAs multi-quantum-disk NW | 10 | 2150 | 1573 | 1311 | RT | 26 |
| InGaAs/InGaP core/shell NW array | 3.5 | 100 | 1100 | - | RT | 32 |
| GaAs/GaAsSb multiple superlattices NW | 10 | 75 | 890-990 | 1250 | RT | 20 |
| AlGaAs/GaAs Multiple Quantum Well NW | 5 | 110 | 777/791/805/819 | 791 | RT | 33 |
| Zn doped GaAs NW | 5.15 | 165 | 867/881/895 | 345 | RT | 34 |
| GaAs/InGaAs/AlGaAs Multiple quantum dots NW | 4.3 | 179 | 879/898 | - | RT | 31 |
| GaAs/InGaAs | 2.2 | 1.6 | 959 | 6400 | 5 K | this work |
| GaAs/InGaAs/AlGaAs | 2.6 | 48 | 945.6 | 1891 | RT | this work |

An ultralow threshold power of ~ 48 μJ/cm$^2$ per pulse and single mode lasing can be achieved at room temperature. Table 1 shows the comparison of cavity length, threshold, lasing peak and *Q-factor* of various NW lasers reported in the literature. The cavity length in this work is the smallest reported which can sustain lasing. The *Q-factor* and lasing threshold are among the best as a result of the excellent NW morphology and the high gain provided by the QDs, making our NWs an excellent single-mode NW laser. Furthermore, by replacing the focus collimation packages with a silicon photodiode in the signal collection path (see Figure S3), the lasing power of ~ 0.9 μW from a single NW is measured under a pump fluence of 59.5 μJ/cm$^2$ per pulse at room temperature. Considering the limitation of the horizontal configuration of the lying NW laser, the actual output power could be higher. The relatively high power output of our NW laser is a major step towards future application.

## CONCLUSIONS

In this work, we demonstrate single-mode lasing from unpassivated GaAs/InGaAs multi-QDs NWs up to 180 K. By designing the cavity size of the NW, the lasing mode can be selected and controlled. The high optical gain provided by the QDs can compensate for the optical loss caused by the reduced cavity length. Furthermore, the excellent NW morphology also results a low cavity loss. Single-mode lasing at room temperature is obtained by passivating the NW surface an AlGaAs/GaAs shell to further reduce non-radiative losses from the surface. Our QDs NW lasers exhibit single-mode output with a very low threshold fluence ~ 48 μJ/cm$^2$ per pulse at room temperature with a high characteristic temperature ~223 K and power output ~ 0.9 μW. This realization of excellent single-mode, ultralow threshold NW lasers are promising for the development of high-powered nano-scale laser for future applications.

**METHODS**

**Growth method:** GaAs/InGaAs QD NWs were grown on GaAs (111)B substrates *via* SA-MOVPE in an AIXTRON 200/4 reactor. Before growth, ~30 nm of SiO$_2$ was first deposited on the GaAs (111)B substrates at 300 °C by plasma-enhanced chemical vapor deposition. Then hexagonal arrays of holes were patterned by electron beam lithography, followed by reactive ion etching. After trim etching with sulfuric acid and hydrochloric acid, the substrate was loaded into the growth chamber. Trimethylgallium, trimethylindium and arsine were used as precursors for the Ga, In and As, respectively. The details of the growth conditions can be found in the supporting information.

**Optical experiment method:** A confocal photoluminescence microscopy system (shown in Figure S3) was used for the optical characterization of the NWs. An 800 nm femtosecond laser (repetition rate: 85 MHz, pulse width: 35 fs) was used to excite the NWs. Two ND filters were used to attenuate the intensity of the pump laser and the emission from the NWs to prevent saturation of the CCD camera attached to the spectrometer. A Lp filter was used to eliminate scattering of pump laser from collected PL signals. A pair of lenses (objective lens and the lens in FCP) and a fiber made up the confocal system, which is capable of pumping the NW and collecting the signals. A spectrometer was used for emission spectrum analysis by using the attached CCD 2, and the carrier lifetime measurement was performed using a time-corelated single photon counting (TCSPC) system composed of the attached single photon detector (SPD, resolution: 50 ps) and a Multi-Channel Scaling (MCS) board (resolution: 25 ps). A HWP and LP package was used for polarization dependence measurements of the emission spectrum. A flip mirror was used for switching between spectrum analysis and PL imaging, which was achieved by a lens and CCD 1. For low temperature measurements, a cryostat operating in the range of 4 to 300 K was used.

## ASSOCIATED CONTENT

**Supporting Information**

The Supporting Information is available free of charge on the ACS Publications website.

Normalized PL spectral mapping of the InGaAs QDs; energy-dispersive spectroscopy (EDS) analysis; schematic diagram of experiment setup; transverse mode simulations; lasing power output measurement; emission spectra of a GaAs/InGaAs multi-QDs NW; lasing characterization from two different GaAs/InGaAs multi-QDs NWs; the spontaneous emission spectrums under low pump fluence; lasing characterization from a GaAs/InGaAs multi-QDs NW at 180 K; emission spectra from a GaAs/InGaAs/AlGaAs multi-QDs NW at different temperature; simulations of polarization dependence; mode hopping analysis; optical image from a GaAs/InGaAs/AlGaAs multi-QDs NW; growth conditions.

## AUTHOR INFORMATION

**Corresponding Authors:**

*E-mail: xuetaogan@nwpu.edu.cn

*E-mail: ppchen@mail.sitp.ac.cn

*E-mail: luwei@shanghaitech.edu.cn

**ORCID**

Xutao Zhang: 0000-0001-5259-5745

Ziyuan Li: 0000-0001-9400-6902

Jianlin Zhao: 0000-0002-4619-1215

Lan Fu: 0000-0002-9070-8373

Hark Hoe Tan: 0000-0002-7816-537X

**Notes**

The authors declare no competing financial interest.

**ACKNOWLEDGMENTS**

X. Z. and R. Y. contributed equally to this work. This work is supported by the National Natural Science Foundation of China (62005222, 61974166, 12027805, 11991060, 61775183, 91950119, 61905196), the Key Research and Development Program (2017YFA0303800), the Key Research and Development Program in Shaanxi Province of China (2020JZ-10), the Natural Science Basic Research Program of Shaanxi Province (2020JQ-222), the Shanghai Science and Technology Committee (18JC1420401), the China Postdoctoral Science Foundation (2020M683548) and the Australian Research Council. The Australian National Fabrication Facility is acknowledged for access to the epitaxial growth facilities.

# Supporting Information

# Ultralow Threshold, Single-Mode InGaAs/GaAs Multi-Quantum Disk Nanowire Lasers

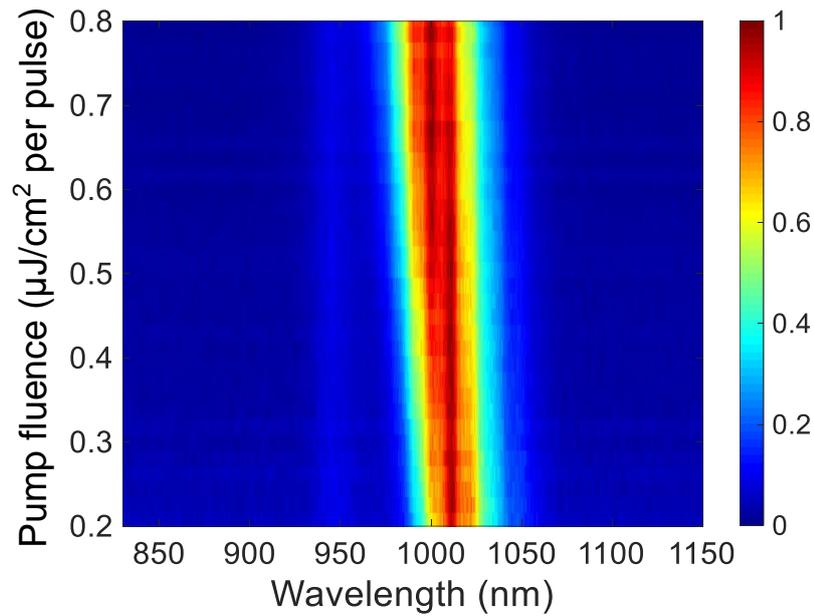

Figure S1. Normalized PL spectral as a function of the pump fluence from a NW, showing the InGaAs QDs emitting around 1000 nm with a linewidth of ~50 nm.

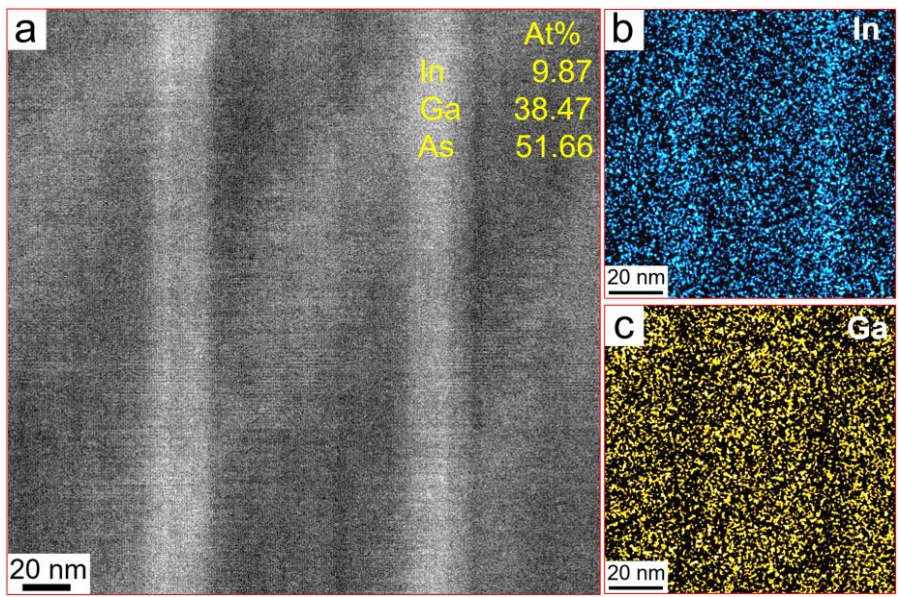

Figure S2. (a) Scanning transmission electron microscopy image of the QDs region. (b-c) Corresponding EDS Ga and In maps, confirming the InGaAs QDs has a In/Ga/As ratio of 9.87:38.47:51.66.

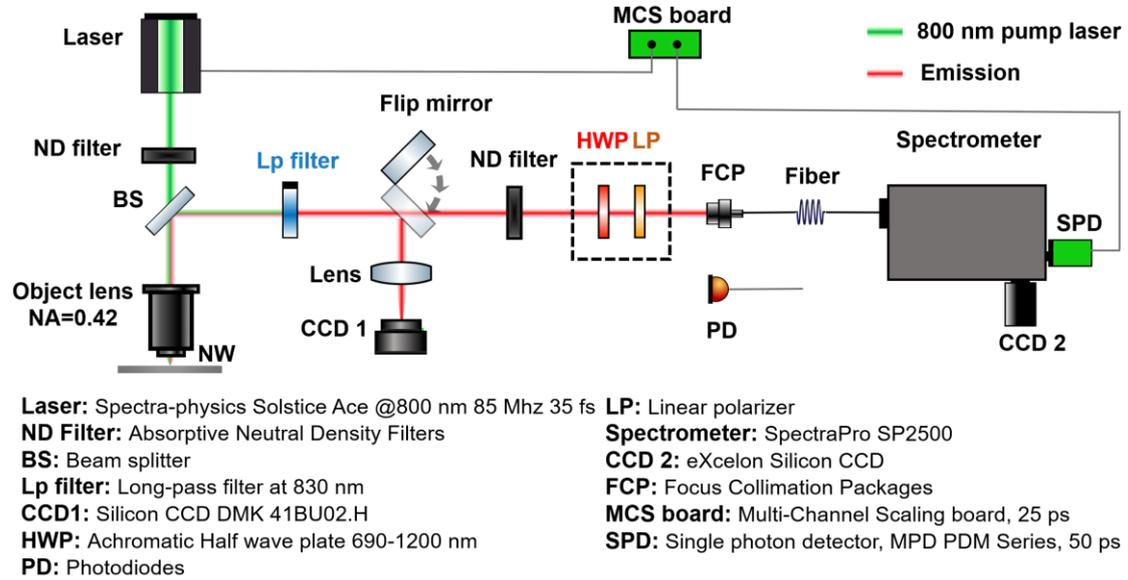

Figure S3. Schematic diagram of experiment setup used for optical characterization, including pump fluence and polarization dependent emission spectrum, photoluminescence imaging, and carrier lifetime measurements.

**Transverse mode simulations:** The two dimensional (2D) finite difference eigenmode (FDE) solver was used to calculate the supported modes in the NW. A model with a GaAs NW lying on the $SiO_2$ substrate was used. A square shaped 2D FDE solver region was set at the center of the cross-section of NW (Size: 5 times of NW diameter; Boundary setting: Metal). Only modes with effective index above 1.45 (refractive index of $SiO_2$) could be considered and shown in the calculation result.

**Lasing power output measurement:** By replacing the FCP in Figure S3 with a silicon photodiode, the lasing power of ~ 31.5 nW is measured in the signal collection path from a single GaAs/InGaAs/AlGaAs multi-QDs NW under a pump fluence of 59.5 µJ/cm² per pulse at room temperature. Considering the losses caused by the limited numerical aperture of the objective lens and beam splitters in the experiment setup, the power output is estimated up to be ~ 0.9 µW.

**Loss description:** There are two beam splitters in the actual experiment setup, each of these will introduce a 50 % loss. Considering the NA 0.42 of the objective lens, a 3D FDTD simulation is performed and the ratio between fraction of power collected by objective lens and facet emitting power is 14 %. Losses introduced by other components, such as planar mirrors, Lp filter, are not taken into consideration.

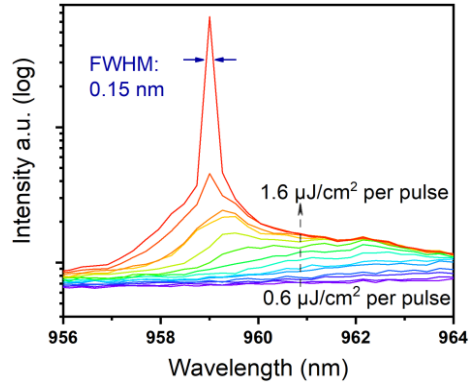

Figure S4. Emission spectra at different pump fluences, showing a linewidth of 0.15 nm at 5 K.

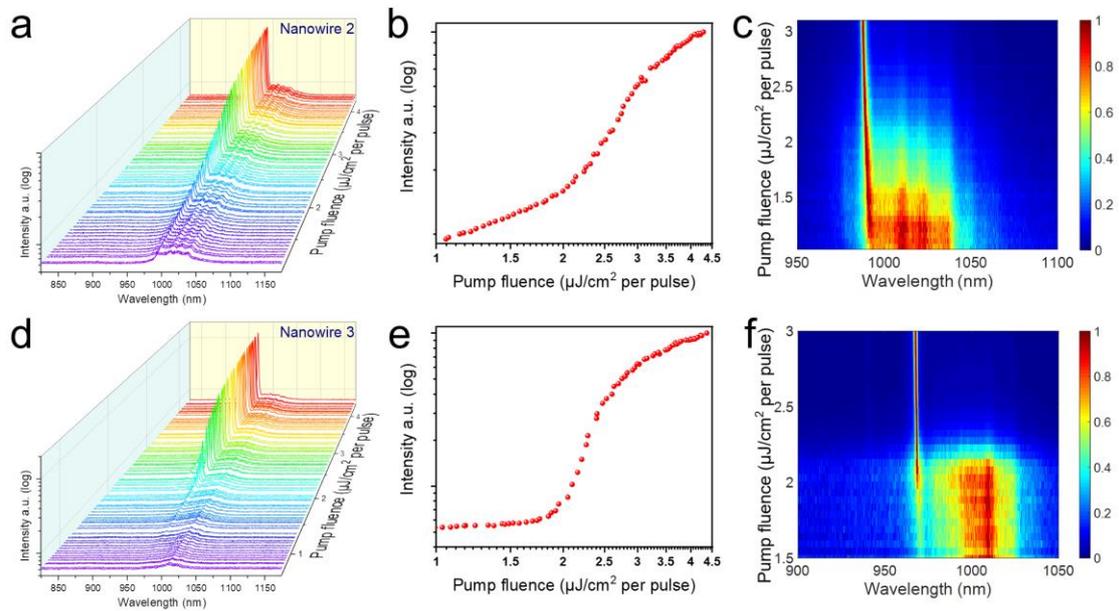

Figure S5. Lasing characterization from two different nanowires (a-c) and (d-f) at 5 K.

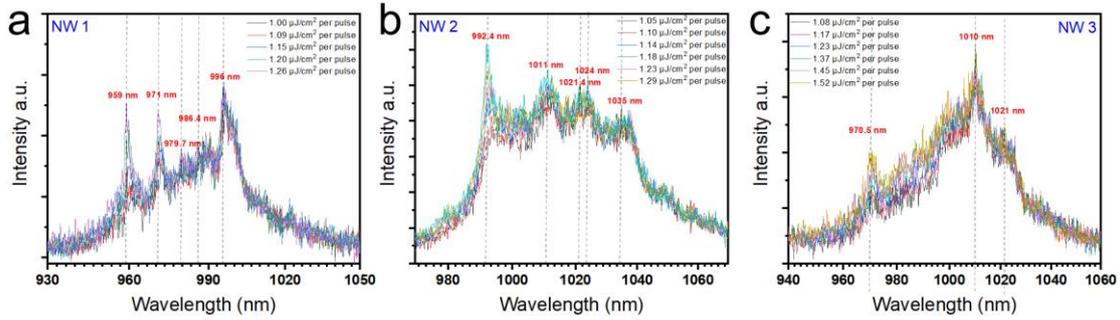

Figure S6. The photoluminescence spectra of the spontaneous emission from different nanowires at low pump fluence. The sharp emission lines in the spectra may be caused by the inhomogeneous GaAs/InGaAs interface characteristics and random InGaAs composition fluctuations.

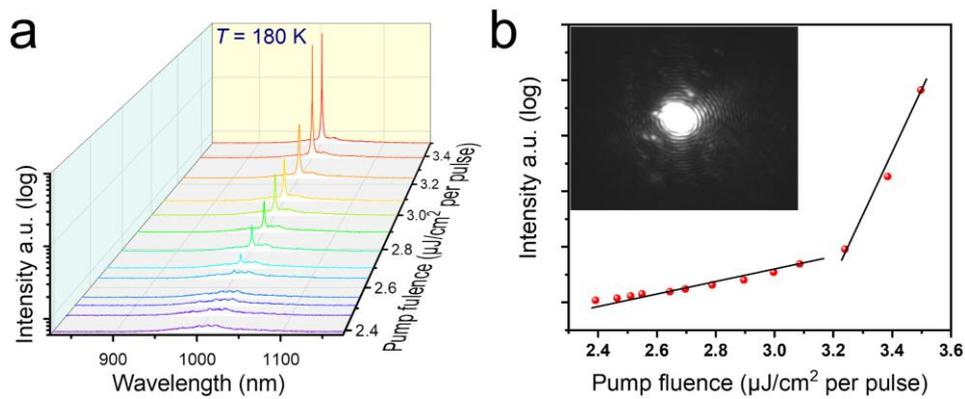

Figure S7. Lasing characterization from a GaAs/InGaAs multi-QDs NW at 180 K. (a) Emission spectra at different pump fluence. (b) Lasing emission intensity as a function of pump fluence. The inset in the upper right corner is the optical image from the NW at lasing.

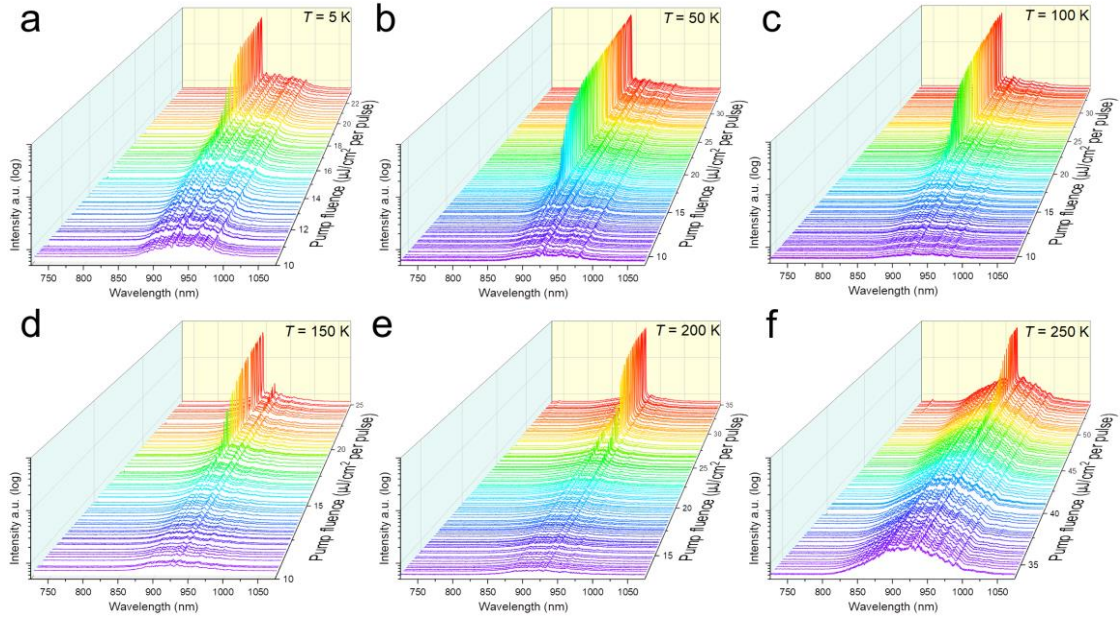

Figure S8. Emission spectra at different pump fluences from a GaAs/InGaAs/AlGaAs multi-QDs NW at 5 K (a), 50 K (b), 100 K (c), 150 K (d), 200 K (e) and 250 K (f).

**Simulations of Polarization dependence:** 3D FDTD simulations were performed to identify the lasing mode of the NW. To simplify the model a GaAs NW lying on a $SiO_2$ substrate was used. A 2D field monitor was placed at a height of 200 nm above the top surface of the NW to record the near field intensity. Far field data was calculated from the near field data. Considering the limited numerical aperture (NA) of the objective lens, only a fraction of electromagnetic wave will be collected by the experimental setup. Since we used an objective lens with NA=0.42, the far field intensity for different polarization angle, relative to the nanowire axis, was calculated first, and then integrated from 0 to 24.85° for polar angle in spherical coordinates. In this way, we obtained the polarization dependence for each guided mode.

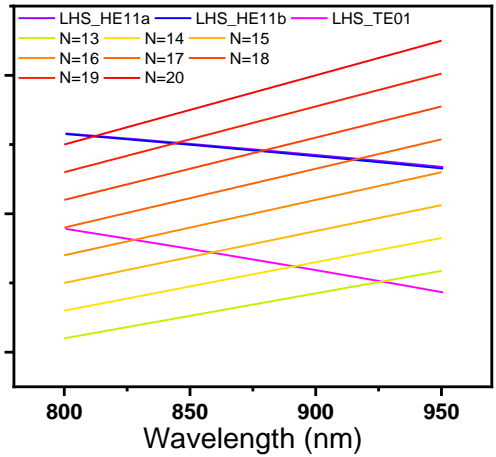

Figure S9. Graphical solutions of $2Ln_{eff}=N\lambda$ to determine the lasing mode of the NW. Left hand side (LHS) (HE$_{11a}$/HE$_{11b}$ and TE$_{01}$) and right hand side (RHS) of $2Ln_{eff} = N\lambda$ are plotted *versus* wavelength. The intersection points, marked by filled circles, determine the lasing mode and wavelength.

**Mode hopping analysis:** The mode spacing between the new and the previous peak is about 25 nm, which is different from the HE$_{11a}$/HE$_{11b}$ Fabry–Pérot mode spacing for this NW, calculated to be 40 nm. Considering the passivation cap/shell resulting in an increase of overall NW diameter to ~300 nm, TE$_{01}$ mode transverse mode is also supported in the NW. So, the mode might hop between TE$_{01}$ and HE$_{11a}$/HE$_{11b}$, with increasing temperature. By solving the equation $2Ln_{eff}=N\lambda$ for TE$_{01}$ and HE$_{11a}$/HE$_{11b}$, a mode spacing of 25.5 nm between TE$_{01}$ at and HE$_{11a}$ is obtained, which matches well with the experimental mode spacing. The detailed simulation is discussed as follows: The lasing mode and wavelength are determined by:

$$2Ln_{eff} = N\lambda \qquad (1)$$

$n_{eff}$ *versus* wavelength for different transverse mode is obtained by Lumerical MODE simulation, for a GaAs NW (diameter 300 nm and length 2.6 μm) lying on a SiO$_2$ substrate. Through a graphical solution of equation (1) shown in Figure S9, two intersections at 916.9 nm (HE$_{11a}$) and 891.4 nm (TE$_{01}$) are pretty close to the experimental lasing peaks of GaAs/InGaAs/AlGaAs NW at 150 K, which are 917.4 and 892.8 nm. Since we use the room temperature value of the refractive index for simulation, the lasing peak and the solutions do not match well under the broad range of temperatures investigated. Optical excitation can also heat up the NW, so actual temperature of NW should be higher than operating temperature, which can also contribute to the slight mismatch between solutions and lasing peak at room temperature.

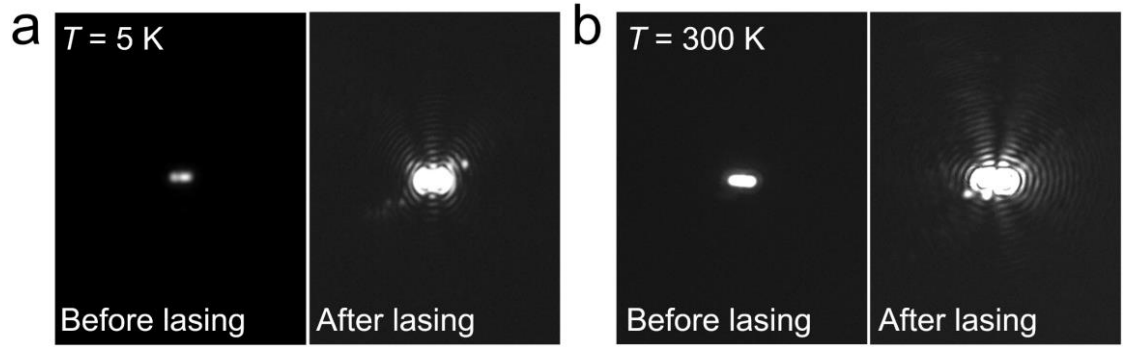

Figure S10. Optical image from single GaAs/InGaAs/AlGaAs multi-QDs NW before and after lasing at 5 K (e) and room temperature (f), respectively.

**Growth conditions:** The optimized growth conditions of the GaAs/InGaAs QD NWs are as follows:

| Growth Step | TMGa (mol/min) | TMIn (mol/min) | AsH$_3$ (mol/min) | TMAl (mol/min) | Time | T (°C) |
|---|---|---|---|---|---|---|
| GaAs base | 1.006E-05 | 0 | 1.998E-04 | 0 | 20 min | 750 |
| InGaAs QD | 1.006E-05 | 2.529E-06 | 1.998E-04 | 0 | 15 s | 750 |
| GaAs Barrier | 1.006E-05 | 0 | 1.998E-04 | 0 | 30 s | 750 |
| GaAs top | 1.006E-05 | 0 | 1.998E-04 | 0 | 10 min | 750 |

The optimized growth conditions of the GaAs/InGaAs QD NWs with AlGaAs shell are as follows:

| Growth Step | TMGa (mol/min) | TMIn (mol/min) | AsH$_3$ (mol/min) | TMAl (mol/min) | Time | T (°C) |
|---|---|---|---|---|---|---|
| GaAs base | 1.006E-05 | 0 | 1.998E-04 | 0 | 20 min | 750 |
| InGaAs QD | 1.006E-05 | 2.529E-06 | 1.998E-04 | 0 | 15 s | 750 |
| GaAs Barrier | 1.006E-05 | 0 | 1.998E-04 | 0 | 30 s | 750 |
| GaAs top | 1.006E-05 | 0 | 1.998E-04 | 0 | 5 min | 750 |
| AlGaAs shell | 6.683E-06 | 0 | 4.855E-04 | 5.464E-06 | 2 min | 750 |
| GaAs cap | 1.006E-05 | 0 | 1.998E-04 | 0 | 2 min | 750 |

Notes: To get straight and regular shape of the GaAs/InGaAs QD nanowires, high-quality mask without any unexpected pinholes is the precondition and a cleaning process is necessary to ensure that the substrate and mask are free of residual organic matter and oxides. Secondly, a higher growth temperature is beneficial for achieving the taper-free morphology and avoiding the accompanying island-like particle growth on the substrate.